\documentclass[aps,pra,preprint,onecolumn,letterpaper]{revtex4}

\usepackage{graphicx}                            
\usepackage{epstopdf}
\usepackage{amsmath} 
\usepackage{amssymb}
\usepackage{quotes}
\usepackage{color}
\usepackage{transparent}
\usepackage{dcolumn}
\usepackage{bm}
\usepackage{multirow}
\usepackage{cancel} 
\usepackage{mdframed}
\usepackage{color}
\usepackage{soul, color} 
\soulregister\ref{7}  
\soulregister\cite{7} 
\renewcommand{\st}[1]{}
\renewcommand{\hl}[1]{#1}

\begin{document}

\title{Controlling Directionality and Dimensionality of Wave Propagation through Separable Bound States in the Continuum}
\author{Nicholas Rivera$^{1}$, Chia Wei Hsu $^{1,2}$, Bo Zhen $^{1,3}$, Hrvoje Buljan$^{4}$, John D. Joannopoulos$^{1}$ \& Marin Solja\v{c}i\'{c}$^{1}$}

\affiliation{$^{1}$Department of Physics, Massachusetts Institute of Technology, Cambridge, MA 02139, USA \\ 
$^{2}$Department of Applied Physics, Yale University, New Haven, CT 06520, USA. \\
$^{3}$Physics Department and Solid State Institute, Technion, Haifa 32000, Israel \\
$^{4}$Department of Physics, University of Zagreb, Zagreb 10000, Croatia.  }

\clearpage
\begin{abstract}
\noindent	

A bound state in the continuum (BIC) is an unusual localized state that is embedded in a continuum of extended states. \st{BICs can arise from the separability of the wave equation.} Here, we present the general condition for \st{separable} BICs to \st{exist, which allows us  to control} \hl{arise from wave equation separability and show that} the directionality and dimensionality of \hl{their} resonant radiation \hl{can be controlled} by exploiting \hl{perturbations of certain} symmetry.
Using this general framework\st{ for separable BICs}, we construct new examples of \st{them} \hl{separable BICs} in realistic models of optical potentials for ultracold atoms, photonic systems, and systems described by tight binding. Such BICs with easily reconfigurable radiation patterns allow for applications such as the storage and release of waves at a controllable rate and direction, systems that switch between different dimensions of confinement, and experimental realizations in atomic, optical, and electronic systems. 
\end{abstract}

\maketitle
 \section{Introduction}
The usual frequency spectrum of waves in an inhomogeneous medium consists of localized waves whose frequencies lie outside a continuum of delocalized propagating waves. For some rare systems, the bound states can be at the same frequency as an extended state. \st{These states, called} \hl{Such} bound states in the continuum (BICs) represent a method of wave localization that does not forbid outgoing waves from propagating, unlike traditional methods of localizing waves such as potential wells in quantum mechanics, conducting mirrors in optics, band-gaps, and Anderson localization.

BICs were first \st{shown to exist} \hl{predicted} theoretically in quantum mechanics by von Neumann and Wigner~\cite{neumann1929}. However, their BIC-supporting potential was highly oscillatory and could not be implemented in reality. The BIC was seen as a mathematical curiosity for the following 50 years until Friedrich and Wintgen showed that BICs could exist in models of multi-electron atoms~\cite{friedrich1985}. Since then, many theoretical examples of BICs have been demonstrated in quantum mechanics~\cite{robnik1986,nockel1992,duclos2001,prodanovic2013, ladron2003,ordonez2006,moiseyev2009}, 
 electromagnetism~\cite{ochiai2001,ctyroky2001,watts2001,kawakami2002,apalkov2003,shipman2005,porter2005,bulgakov2008,marinica2008,molina2012,2013_Hsu_LSA,bulgakov2014,monticone2014,2014_Bulgakov_OL}, \hl{acoustics~\cite{porter2005}, and water waves~\cite{cobelli2011},} \st{and more general wave frameworks. On the other hand, experimental realizations are much fewer in number} 
\hl{with some experimentally realized~\cite{plotnik2011,lee2012,2013_Hsu_Nature,corrielli2013,weimann2013,lepetit2014}.}
A few mechanisms explain most examples of BICs that have been discovered. \st{One of the} \hl{The} simplest mechanism is \st{that of the} symmetry\st{-protected BICs} \hl{ protection}, in which the BIC is decoupled from its degenerate continuua by symmetry mismatch~\cite{ochiai2001,fan2002,ladron2003,plotnik2011,lee2012,shipman2005,moiseyev2009}. \st{Another type arises when two resonances couple, leading one of them to become a BIC \cite{friedrich1985,marinica2008,lepetit2014}. When two resonant states with non-zero imaginary part of their frequency cross in the real part of their frequency as a parameter in the Hamiltonian is varied, one of these resonances can become a BIC.}\hl{Or, when two resonances are coupled, they can form BICs through Fabry-Perot interference~\cite{ordonez2006,marinica2008, 2013_Hsu_LSA, weimann2013} or through radiative coupling with tuned resonance frequencies~\cite{friedrich1985,lepetit2014,2014_Bulgakov_OL}.} More generally, parametric BICs can occur for special values of  parameters in the Hamiltonian~\cite{porter2005,bulgakov2008,2013_Hsu_Nature,yi2014,bulgakov2014,monticone2014}, which can in some cases be understood using topological arguments~\cite{zhen2014}.

There exists yet another class of BICs, which has thus far been barely explored, and literature covering it is rare and scattered; these are BICs \hl{due to separability}~\cite{robnik1986,nockel1992,prodanovic2013,ctyroky2001,watts2001,kawakami2002,apalkov2003,duclos2001}. In this paper, we develop  the most general properties of separable BICs \st{-} by providing the general criteria for \st{the} \hl{their} existence \st{of separable BICs} and by characterizing the continuua of these BICs. We demonstrate that separable BICs enable control over both the directionality and the dimensionality of resonantly emitted waves by exploiting symmetry, which is not possible in other classes of BICs. Our findings \hl{may} lead to applications such as\st{: a new class of} tunable-$Q$ directional resonators and quantum systems which can be switched between quantum dot (0D), quantum wire (1D), and quantum well (2D) modes of operation. While previous works on separable BICs have been confined to somewhat artificial systems, we \hl{also} present readily experimentally observable examples of separable BICs in photonic systems with directionally resonant states, as well as examples in optical potentials for cold atoms. 

\section{Results}
\subsection{General Condition for Separable BICs}

We start by considering a simple example of a two-dimensional separable system - one where the Hamiltonian operator can be written as a sum of Hamiltonians that each act on distinct variables $x$ or $y$, i.e;
\begin{equation}
H = H_x(x) + H_y(y).
\end{equation}
Denoting the eigenstates of $H_x$ and $H_y$ as $\psi_{x}^i(x)$ and $\psi_{y}^j(y)$, with energies $E_{x}^i$ and $E_{y}^j$, it follows that $H$ is diagonalized by the basis of product states $\psi_{x}^i(x)\psi_{y}^j(y)$ with energies $E^{i,j}=E_{x}^i+E_{y}^j$. If $H_x$ and $H_y$ each have a continuum of extended states starting at zero energy, and these Hamiltonians each have at least one bound state, $\psi_x^0$ and $\psi_y^0$, respectively, then the continuum of $H$ starts at energy $\text{min}(E_{x}^0,E_{y}^0)$, where the $0$ subscript denotes ground states. Therefore, if there exists a bound state satisfying  $E^{i,j}>\text{min}(E_{x}^0,E_{y}^0)$, then it is a bound state in the continuum. We schematically illustrate this condition being satisfied in Figure 1, where we illustrate a separable system in which $H_x$ has two bound states, $\psi_x^0$ and $\psi_x^1$, and $H_y$ has one bound state, $\psi_y^0$. The first excited state of $H_x$ combined with the ground state of $H_y$, $\psi_x^1\psi_y^0$, while spatially bounded, has larger energy, $E_x^1 + E_y^0$, than the lowest continuum energy, $E_x^0$, and is therefore a BIC. 

To extend separability to a Hamiltonian with a larger number of separable degrees of freedom is then straightforward. The Hamiltonian can be expressed in tensor product notation as $H =  \sum\limits_{i=1}^N H_i$, where $H_i \equiv I^{\otimes i-1}\otimes h_i \otimes I^{\otimes N-i}$. In this expression, $N$ is the number of separated degrees of freedom, $h_i$ is the operator acting on the $i$-th variable, and $I$ is the identity operator. The variables may refer to the different particles in a non-interacting multi-particle system, or the spatial and polarization degrees of freedom of a single-particle system. Denote the $n_j$th eigenstate of $h_i$ by $|\psi_{i}^{n_j}\rangle$ with energy $E_i^{n_j}$. Then, the overall Hamiltonian $H$ is trivially diagonalized by the product states $|n_1,n_2,\cdot\cdot\cdot ,n_N\rangle \equiv |\psi_1^{n_1}\rangle\otimes|\psi_2^{n_2}\rangle\otimes\cdot\cdot\cdot\otimes|\psi_N^{n_N}\rangle$ with corresponding energies $E = E_1^{n_1} + E_2^{n_2} + \cdot\cdot\cdot + E_N^{n_N}$. Denoting the ground state of $h_i$ by $E_i^{0}$ and defining the zeros of the $h_i$ such that their continuua of extended states start at energy zero, the continuum of the overall Hamiltonian starts at $\underset{{1\leq i \leq N}}{\text{min}}(E_{i}^{0})$. Then, if the separated operators $\{ h_i \}$ are such that there exists a combination of separated bound states satisfying $E = \sum\limits_{i=1}^N E_{i}^{n_i} > \underset{{1\leq i \leq N}}{\text{min}}(E_{i}^{0})$, this combined bound state is a BIC of the overall Hamiltonian. For such separable BICs, coupling to the continuum is forbidden by the separability of the Hamiltonian. Stating the most general criteria for separable BICs allows us to straightforwardly extend the known separable BICs to systems in three dimensions, multi-particle systems, and systems described by the tight-binding approximation.

\subsection{Properties of the Degenerate Continuua}

A unique property that holds for all separable BICs is that the delocalized modes degenerate to the BIC are always guided in at least one direction. In many cases, there are multiple degenerate delocalized modes and they are guided in different directions. In 2D, we can associate partial widths $\Gamma_x$ and $\Gamma_y$ to the resulting resonances of these systems when the separable system is perturbed. These partial widths are the decay rates associated with energy-conserving transitions between the BIC and states delocalized in the $x$ and $y$ directions under perturbation, respectively. When separability is broken, we generally can not decouple leakage in the $x$ and $y$ directions because the purely-$x$ and purely-$y$ delocalized continuum states mix. However, in this section, we show that one can control the radiation to be towards the x (or y) directions only, by exploiting the symmetry of the perturbation.

In Fig. 2(a), we show a two-dimensional potential in a Schrodinger-like equation which supports a separable BIC. This potential is a sum of Gaussian wells in the $x$ and $y$-directions, given by \begin{equation} V(x,y) = -V_xe^{-\frac{2x^2}{\sigma_x^2}}-V_ye^{-\frac{2y^2}{\sigma_y^2}}. \end{equation} This type of potential is representative of realistic optical potentials for ultracold atoms and potentials in photonic systems. Solving the time-independent Schrodinger equation for $\{V_x, V_y \}=\{1.4, 2.2\}$ and $\{\sigma_x,\sigma_y \} = \{5,4\}$ (in arbitrary units) gives the spectra shown in Fig. 2(b).
Due to the $x$ and $y$ mirror symmetries of the system, the modes have either even or odd parity in both the $x$ and the $y$ directions.
This system has several BICs. Here we focus on the BIC $|n_x,n_y\rangle = |2,1\rangle$ at energy $E^{2,1} =-1.04$, with the mode profile shown in Fig.~2(c); being the second excited state in $x$ and the first excited state in $y$, this BIC is even in $x$ and odd in $y$. It is only degenerate to continuum modes $|0,E^{2,1}-E_{x}^0\rangle$ extended in the $y$ direction (Fig. 2(d)), and $|E^{2,1}-E_{y}^0,0\rangle$ extended in the $x$ direction (Fig. 2(e)), where an $E$ label inside a ket denotes an extended state with energy $E$. 

If we choose a perturbation that preserves the mirror symmetry in the $y$-direction, as shown in the inset of Fig.~2(f), then the perturbed system still exhibits mirror symmetry in $y$. Since the BIC $|2,1\rangle$ is odd in $y$ and yet the $x$-delocalized continuum states $|E^{2,1}-E_{y}^0,0\rangle$ are even in $y$, there is no coupling between the two. As a result, the perturbed state radiates only in the $y$ direction, as shown in the calculated mode profile in Fig.~2(f).
This directional coupling is a result of symmetry, and so it holds for arbitrary perturbation strengths.

On the other hand, if we apply a perturbation that is odd in $x$ but not even in $y$, as shown in the inset of Fig.~2(g), then there is radiation in the $x$ direction only  to first-order in time-dependent perturbation theory.
Specifically, for weak perturbations of the Hamiltonian, $\delta V$, the first-order leakage rate is given by Fermi's Golden Rule for bound-to-continuum coupling, $\Gamma \sim \sum_c |\langle 2,1 |\delta V| \psi_c \rangle|^2\rho_c(E^{2,1})$, where $\rho_c(E)$ is the density of states of continuum $c$, and $c$ labels the distinct continuua which have states at the same energy as the BIC.
Since the BIC and the $y$-delocalized continuum states $|0,E^{2,1}-E_{x}^0\rangle$ are both even in $x$, the odd-in-$x$ perturbation does not couple the two modes directly, and $\Gamma_y$ is zero to the first order. As a result, the perturbed state radiates only in the $x$ direction, as shown in Fig.~2(g). At the second order in time-dependent perturbation theory, the BIC can make transitions to intermediate states $k$ at any energy, and thus the second-order transition rate, proportional to  $\sum_c |T_{ci}|^2 = \sum_c |\sum_k \frac{\langle c|\delta V|k \rangle\langle k|\delta V|i\rangle}{E_i - E_k + i0^+}|^2$ does not vanish because the intermediate state can have even parity in the $x$-direction. 

Another unique aspect of separability is that by using separable BICs in 3D, the dimensionality of the confinement of a wave can be switched between one, two, and three by tailoring perturbations applied to a single BIC mode. The ability to do this allows for a device which can simultaneously act as a quantum well, a quantum wire, and a quantum dot. We demonstrate this degree of control using a separable potential generated by the sum of three Gaussian wells (in $x$, $y$, and $z$ directions) of the form in Equation 2 with strengths $\{0.4,0.4,1 \}$ and widths $\{ 12,12,3 \}$, all in arbitrary units. The identical $x$ and $y$ potentials have four bound states at energies $E_x = E_y = -0.33, -0.20, -0.10 \text{ and } -0.029$. The $z$ potential has two bound states at energies $E_z =-0.61 \text{ and } -0.059$. All of the continuua start at zero. The BIC state $|1,1,1\rangle$ at energy $E^{1,1,1}=-0.47$ is degenerate to $|0,0,E_z\rangle,|0,1,E_z\rangle,|1,0,E_z\rangle,|E_x,m,0\rangle,|n,E_y,0\rangle$ and $|E_x,E_y,0\rangle$, where $E_i$ denotes an energy above zero, and $m$ and $n$ denote bound states of the $x$ and $y$ wells. For perturbations which are even in the $z$ direction, the BIC $|1,1,1\rangle$ does not couple to states delocalized in $x$ and $y$ ($|E_x,m,0\rangle,|n,E_y,0\rangle,\text{ and }|E_x,E_y,0\rangle$), because they have opposite parity in $z$, so the BIC radiates in the $z$-direction only. On the other hand, for perturbations which are even in the $x$ and $y$ directions, the coupling to states delocalized in $z$ ($|0,0,E_z\rangle,|0,1,E_z\rangle,\text{ and }|1,0,E_z\rangle$) vanishes, meaning that the BIC radiates in the $xy$-plane. 

\subsection{Proposals for the experimental realizations of separable BICs}

BICs are generally difficult to experimentally realize because they are fragile under perturbations of system parameters. On the other hand, separable BICs are straightforward to realize and also robust with respect to changes in parameters that preserve the separability of the system. This reduces the difficulty of experimentally realizing separable BICs to ensuring separability. And while this is still in general nontrivial, it can be straightforwardly achieved in systems that respond linearly to the intensity of electromagnetic fields. In the next two realistic examples of BICs that we demonstrate, we use detuned light sheets to generate separable potentials in photonic systems and also for ultracold atoms. 

\textit{Paraxial optical systems $-$} As a first example of this, consider electromagnetic waves propagating paraxially along the 
$z$-direction, in an optical medium with spatially non-uniform index 
of refraction $n(x,y) = n_{0} + \delta n(x,y)$, where $n_0$ is the constant 
background index of refraction and $\delta n \ll n_{0}$. 
The slowly varying amplitude of the electric field $\psi(x,y,z)$ satisfies the 
two-dimensional Schr\"{o}dinger equation (see Ref. 34 and references therein),

\begin{equation}
i\frac{\partial \psi}{\partial z}=
-\frac{1}{2k}\nabla_{\perp}^2 \psi-\frac{k \delta n }{n_0}\psi.
\label{paraxial}
\end{equation}

Here, $\nabla_{\perp}^2=\partial^2/\partial x^2+\partial^2/\partial y^2$, and 
$k=2\pi n_0/\lambda$, where $\lambda$ is the wavelength in vacuum.
The modes of the potential $\delta n(x,y)$ are of the form $\psi=A_j(x,y)e^{i\beta_j z}$, 
where $A_j$ is the profile, and $\beta_j$ the propagation constant of the 
$j$th mode.  
In the simulations we use $n_0=2.3$, and $\lambda=485$~nm.

Experimentally, there are several ways of producing potentials  $\delta n(x,y)$ 
of different type, e.g., periodic\cite{fleischer2005}, 
random\cite{schwartz2007}, and quasicrystal\cite{freedman2006}. 
One of the very useful techniques is the so-called optical induction technique 
where the potential is generated in a photosensitive material (e.g., photorefractives) 
by employing light~\cite{fleischer2003}. 
We consider here a potential generated by two perpendicular light sheets, which 
are slightly detuned in frequency, such that the time-averaged interference vanishes 
and the total intensity is the sum of the intensities of the individual light sheets.
The light sheets are much narrower in one dimension ($x$ for one, $y$ for the other) than the 
other two, and therefore each light sheet can be approximated as having an intensity that depends 
only on one coordinate, making the index contrast separable. 
This is schematically illustrated in Fig. 3(a).

If the sheets are Gaussian along the narrow dimension, then  
the potential is of the form $\delta n(x,y)=\delta n_0 [\exp(-2(x/\sigma)^2)+\exp(-2(y/\sigma)^2)]$. 
It is reasonable to use $\sigma=30$~$\mu$m and $\delta n_0 = 5.7 \times 10^{-4}$. For these parameters, the one dimensional Gaussian wells 
have four bound states, with $\beta$ values of (in mm$^{-1}$): $2.1,1.3,0.55,\text{ and } 0.11$. 
There are eight BICs: $|1,2\rangle,|2,1\rangle,|1,3\rangle,|3,1\rangle$,
$|2,3\rangle$, $|3,2\rangle, |2,2\rangle$, and $|3,3\rangle$. 
Among them, $|1,3\rangle$ and $|3,1\rangle$ are symmetry protected. 
Additionally, the BICs $|1,3\rangle$ and $|3,1\rangle$ can be used to demonstrate directional 
resonance in the $x$ or $y$ directions, respectively, by applying perturbations even in 
$x$ or $y$, respectively. Therefore, this photonic system serves as a platform to demonstrate both separable BICs and directional resonances.

\textit{Optical potentials for ultracold atoms $-$} The next example that we consider can serve as a platform for the first 
experimental measurement of BICs in quantum mechanics. Consider a non-interacting 
neutral Bose gas in an optical potential. Optical potentials are created by employing 
light sufficiently detuned from the resonance frequencies of the atom, where 
the scattering due to spontaneous emission can be neglected, and the atoms are 
approximated as moving in a conservative potential. 
The macroscopic wavefunction of the system is then determined by solving the 
Schr\"{o}dinger equation with a potential that is proportional to the intensity of the light~\cite{bloch2005}.

As an explicit example, consider an ultracold Bose gas of $^{87}Rb$ atoms. 
An optical potential is made by three Gaussian light sheets with equal intensity $I_1 = I_2 = I_3$, widths 
$20 \text{ }\mu\text{m}$, and wavelengths centered at $\lambda=1064 \text { nm}$~\cite{henderson2009}. This is schematically illustrated in Fig. 3(b).
The intensity is such that this potential has depth equal to ten times the recoil energy 
$E_r = \frac{h^2}{2m\lambda^2}$.  
By solving the Schrodinger equation numerically, we find many BICs;
the continuum energy starts at a reduced energy $\xi_c = \frac{2mEx_0^2}{\hbar^2} = -296.24$, 
where $x_0$ is chosen to be $1 \text{ }\mu m$. The reduced depth of the trap is $-446.93$. Each 
one-dimensional Gaussian supports 138 bound states. There are very many BICs in such a system. For concreteness, an example of one is
$|30,96,96\rangle$, with reduced energy $-146.62$. 

\textit{Tight Binding Models$-$} The final example that we consider here is an extension of the separable BIC formalism to systems which are well-approximated by a tight-binding Hamiltonian that is separable. Consider the following one-dimensional tight-binding Hamiltonian, $H_i$, which models a one-dimensional lattice of non-identical sites: \begin{equation} H_i = \sum\limits_{k}\varepsilon_i^{k} |k\rangle\langle k|+t_i \sum \limits_{\langle lm \rangle}(|l\rangle\langle m| + |m\rangle\langle l| ), \end{equation} where $\langle lm \rangle$ denotes nearest-neighbors,  $\varepsilon_i^k $ is the on-site energy of site $k$, and $k,l,\text{ and }m$ run from $-\infty$ to $\infty$. Suppose $\varepsilon_i^k = -V$ for $|k| < N$, and zero otherwise. For two Hamiltonians of this form, $H_1$ and $H_2$, $H_1 \otimes I + I \otimes H_2$ describes the lattice in Fig. 3(c). If we take $H_1 = H_2$ with $\{ V,t,N \} = \{ -1,-0.3,2 \}$, in arbitrary units, the bound state energies of the 1D-lattices are numerically determined to be $-0.93,-0.74,-0.46,\text{ and } -0.16$.  Therefore the states $|2,2\rangle, |2,3\rangle, |3,2\rangle, |3,3\rangle, |3,1\rangle \text{ and } |1,3\rangle$ are BICs. The last two of these are also symmetry-protected from the continuum as they are odd in $x$ and $y$ while the four degenerate continuum states are always even in at least one direction.  Of course, many different physical systems can be adjusted to approximate the system from Eq. (4), so this opens a path for observing separable BICs in a wide variety of systems.

\section{Summary}

With the general criterion for separable BICs, we have extended the existing handful of examples to a wide variety of wave systems including: three-dimensional quantum mechanics, paraxial optics, and lattice models which can describe 2D waveguide arrays, quantum dot arrays, optical lattices, and solids. These BICs exist in other wave equations such as the 2D Maxwell equations, and the inhomogeneous scalar and vector Helmholtz equations, meaning that their physical domain of existence extends to 2D photonic crystals, microwave optics, and also acoustics.  To our knowledge, they do not exist (except perhaps approximately) in isotropic media in the full 3D Maxwell equations $\nabla \times \nabla \times \bold{E} = \nabla(\nabla \cdot \bold{E}) - \nabla^2\bold{E} = \epsilon \frac{\omega^2}{c^2}\bold{E}$ because the operator $\nabla(\nabla \cdot \text{ })$ couples the different coordinates and renders the equation non-separable. We have demonstrated numerically the existence of separable BICs in models of trapped Bose-Einstein condensates and photonic potentials, showing that separable BICs can be found in realistic systems.  These simple and realistic models can facilitate the observation of BICs in quantum mechanics, which to this date has not been conclusively done.

More importantly, we have demonstrated two new properties unique to separable BICs: the ability to control the direction of emission of the resonance using perturbations, and also the ability to control the dimensionality of the resulting resonance. This may lead to two applications. In the first, perturbations are used as a switch which can couple waves into a cavity, store them, and release them in a fixed direction.   In the second, the number of dimensions of confinement of a wave can be switched between one, two, and three by exploiting perturbation parity. The property of dimensional and directional control of resonant radiation serves as another potential advantage of BICs over traditional methods of localization.

\section{Acknowledgments} The authors would like to acknowledge Prof. Steven G. Johnson, Prof. Marc Kastner, Prof. Silvija Grade\v{c}ak, Wujie Huang, and Dr. Wenlan Chen for useful discussions and advice. Work of M.S. was supported as part of S3TEC, an EFRC funded by the U.S. DOE, under Award Number DE-SC0001299 / DE-FG02-09ER46577.
This work was also supported in part by the MRSEC Program of the National Science Foundation under award number DMR-1419807. This work was also supported in part by the U. S. Army Research Laboratory and the U. S. Army Research Office through the Institute for Soldier Nanotechnologies, under contract number W911NF-13-D-0001.

\section{Author Contributions}  N.R. came up with the idea for separable BICs, and controlling directionality and dimensionality with them, in addition to doing the numerical simulations.  C.W.H., B.Z., H.B., and M.S. helped come up with readily observable physical examples of separable BICs in addition to helping develop the concept of separable BICs. M.S. and J.D.J. provided critical supervision for the work. N.R. wrote the manuscript with critical reading and editing from B.Z., C.W.H., H.B., J.D.J., and M.S..

\bibliographystyle{unsrt}
\bibliography{sepBICsbib}
\clearpage

\begin{figure*}[h]
\begin{centering}
    \includegraphics[width=88mm]{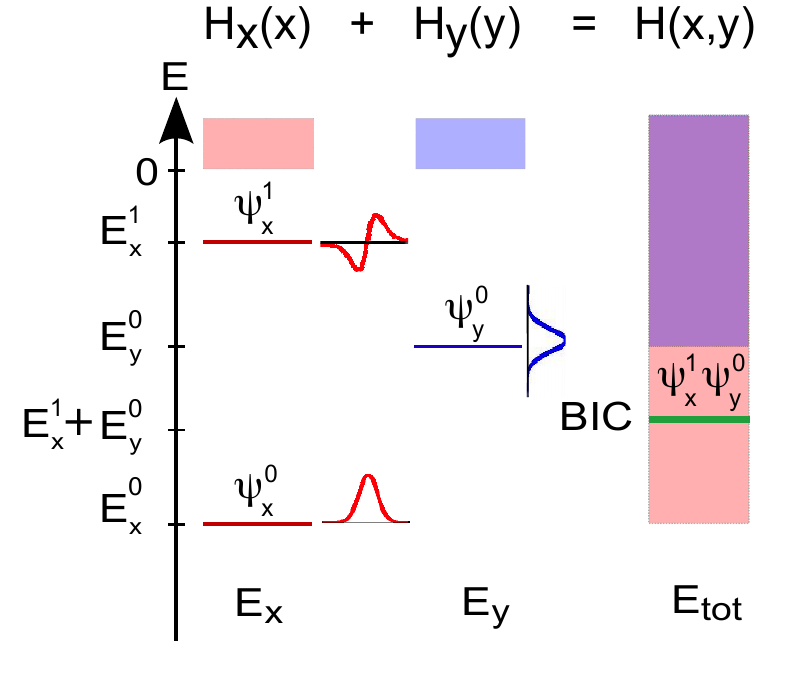}
    \caption{A schematic illustration demonstrating the concept of a separable BIC in two dimensions.}
    \end{centering}
\end{figure*}

\begin{figure*}[h]
  \begin{centering}
    \includegraphics[width=180mm]{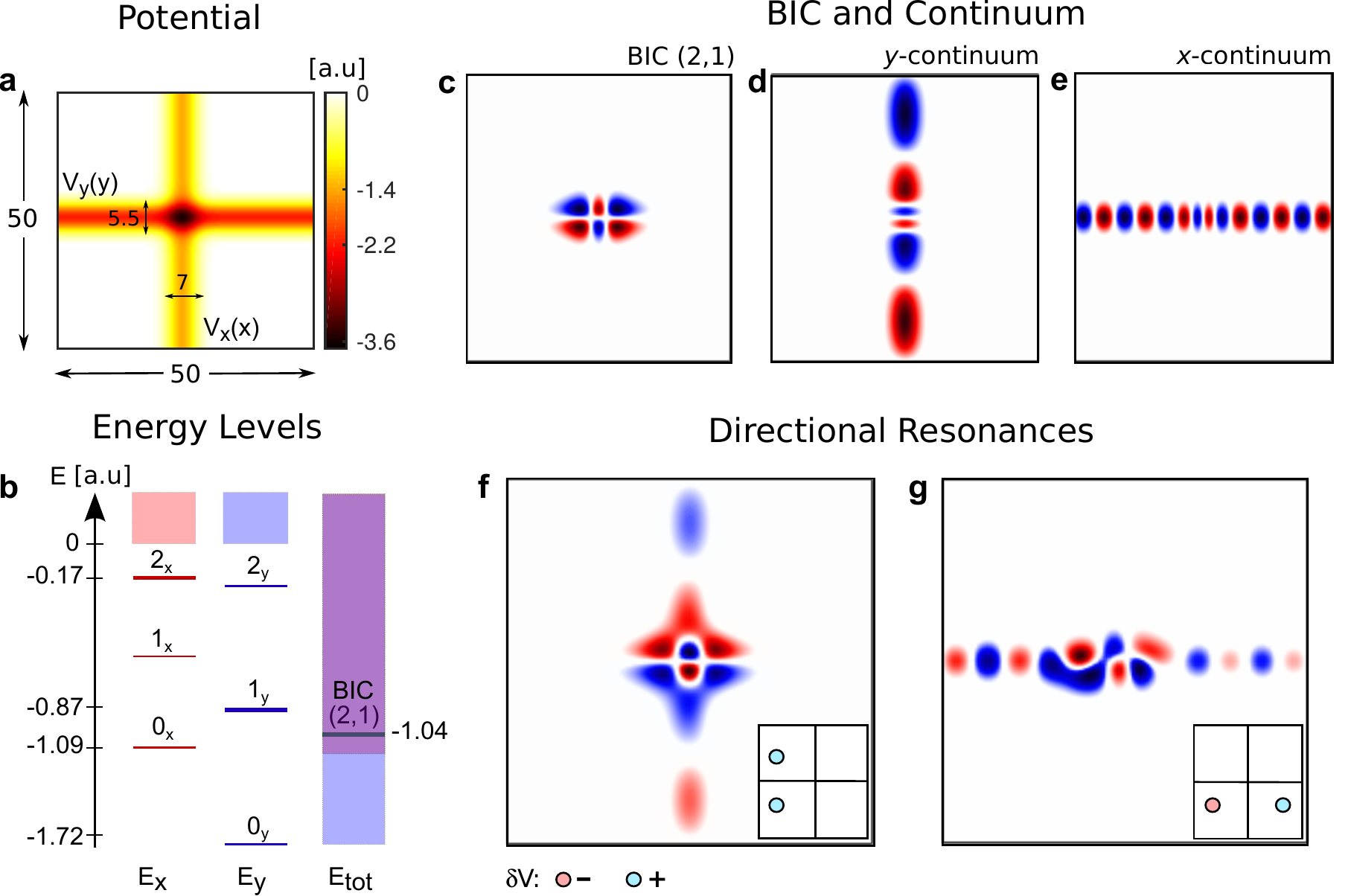}
    \caption{(a) A separable potential which is a sum of a purely x-dependent Gaussian well and a purely y-dependent Gaussian well. (b) The relevant states of the spectrum of the $x$-potential,$y$-potential, and total potential.  (c) A BIC supported by this double well. (d,e) Continuum states degenerate in energy to the BIC. (f) A $y$-delocalized continuum state resulting from an even-$y$-parity perturbation of the BIC supporting potential. (g). An $x$-delocalized continuum state resulting from an odd-$x$-parity perturbation.}
    \end{centering}
\end{figure*}

\begin{figure*}[h]
  \begin{centering}
    \includegraphics[width=180mm]{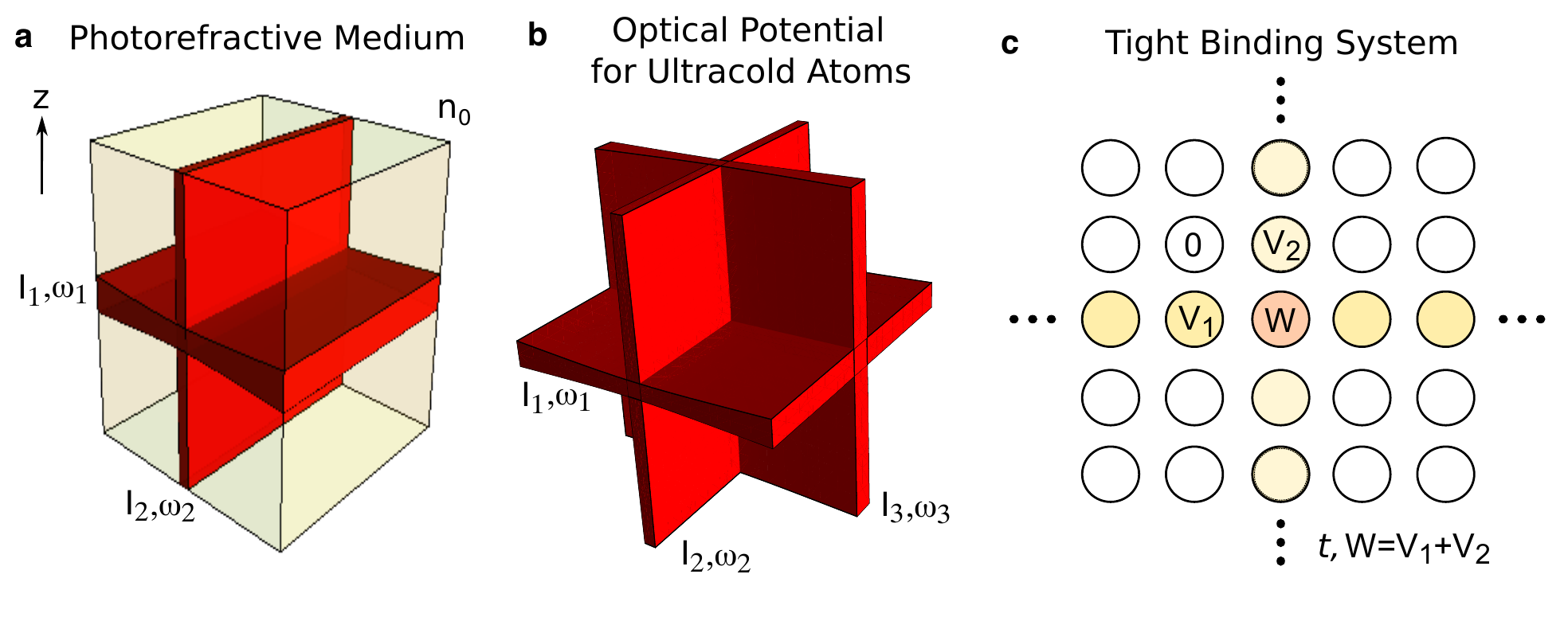}
    \caption{Separable physical systems with BICs. (a) A photorefractive optical crystal whose index is weakly modified by two detuned intersecting light sheets with different intensities. (b) An optical potential formed by the intersection of three slightly detuned light sheets with different intensities. (c) A tight-binding lattice.}
    \end{centering}
\end{figure*}

\end{document}